\documentclass[12pt,a4paper]{article}
\usepackage{amsmath}
\usepackage{amsfonts}
\usepackage{amssymb}
\begin{document}

\author{Andrei Khrennikov\\
International Center for Mathematical Modeling\\
 in Physics and Cognitive Sciences \\
 Linnaeus University,  V\"axj\"o-Kalmar, Sweden}

\title{``Social Laser'': Action Amplification by Stimulated Emission of
Social Energy}
\maketitle

\abstract{The problem of the ``explanation'' of recent social explosions, especially in the Middle East, but also in Southern Europe and the USA, have been 
debated actively in the social and political literature.  We can mention the contributions of  
 P. Mason,  F. Fukuyama,  E. Schmidt and J. Cohen, I. Krastev to this debate.   
We point out that the diversity of opinions and conclusions is really
amazing. At the  moment, there is no consistent and commonly acceptable theory of these phenomena.  
We  present  a model of social explosions  based on a novel approach for the  description of social processes, namely, the  quantum-like approach. 
Here quantum theory is treated simply as an operational formalism - without any direct relation to physics. We explore the quantum-like laser model to 
describe the possibility of Action Amplification by Stimulated Emission of Social Energy (ASE).}

\textbf{keywords} Spontaneous and stimulated absorption and emission, social
energy, Hamiltonian, discrete levels of energy, information excitations

\section{Introduction}

In recent years, we have seen the occurrence of a high level of social
protests throughout the world, see, e.g., \cite{P0}:

{\small ``In the five short years between Occupy Wall Street and Vladimir
Putin's ``Occupy Crimea,'' we witnessed an explosion of protests all around
the world -- the Arab Spring, Russian Winter, Turkish Summer, and the
dismembering of Ukraine all were part of the protest moment. Each of these
demonstrations -- and many less monumental ones -- was angry in its own way,
but the protests are also a worldwide phenomenon.''}

The structure and the causes of this wave of social activation was widely
discussed in a series of publications in the social and political sciences,
see, e.g., \cite{P0}-\cite{P4}. We point out that the diversity of opinions
and conclusions is really amazing. At the moment, there is no consistent and
commonly acceptable theory of these phenomena. In this paper we present a
model of social explosions based on a novel approach used in the description
of social processes, namely the quantum-like approach (see, e.g., the
monographs \cite{QL1}-\cite{QL5} and the references therein). Here quantum
theory is treated \cite{QL1}, \cite{PL1} simply as an operational formalism
- without any direct relation to physics. In this paper we explore the
quantum-like laser model to describe mathematically the possibility of
Action Amplification by Stimulated Emission of Social Energy (ASE). This is
a model of a \textit{social laser}.

In physics the discovery of the laser (light amplification by stimulated
emission of radiation) was based on Einstein's theory of \textit{stimulated
emission of radiation} \cite{EN}. However, it was only in the 1950's that
this theoretical study led to the creation of lasers (1964, the Nobel Prize
went to Charles Hard Townes, Nicolay Gennadiyevich Basov, and Aleksandr
Mikhailovich Prokhorov). Nowadays lasers found numerous applications and can
be considered as one of the main technological outputs of quantum physics.
It could be the case that the model of social laser describing ASE will also
play an important role in the clarification and the description of social
processes and social technologies. This is a pioneer study in this
direction, but a variety of questions have to be clarified in more detail,
see, e.g., section \ref{INT} for a discussion.

We remark that the modern presentation of quantum theory is based on the
advanced mathematical formalism of operator theory in complex Hilbert space.
However, as we know, the pioneering studies of Planck, Einstein, and Bohr
were done before the creation of this mathematical machinery (by Heisenberg,
Schr\"{o}dinger, von Neumann, Dirac). These pioneering studies are known as
\textquotedblleft old quantum mechanics\textquotedblright . Surprisingly,
the most important features of quantum mechanics leading to laser theory
were obtained already in the old quantum mechanics: i.e. the discrete
structure of energy levels for atoms and the quantum structure of
electromagnetic radiation; spontaneous and stimulated emission and
absorption. Here the discrete structure of energy levels of atoms was simply
postulated by Bohr to derive the stability of atoms. Then Einstein
(motivated by Plank's study on black body radiation) postulated the quantum
structure of radiation. By using the quantum structures for atoms and
radiation and thermodynamical considerations, he derived spontaneous and
stimulated emissions and absorption which are fundamental in laser theory.
In our social modeling, such an approach (i.e. a social-information version
of old quantum mechanics) is preferable. Of course, like in the modern
quantum formalism, spontaneous and stimulated radiation processes can be
derived by using the modern theory of open quantum systems. However, the
\textquotedblleft old fashioned considerations\textquotedblright\ in the
spirit of Bohr and Einstein clarify the basic assumptions leading to the
functioning of the laser in a more intuitive and less formal way.

Nowadays the application of physical models outside of physics is well
established and is a rapidly growing research activity. As a non-quantum
example, we can mention \textit{econophysics} \cite{Stanley}, where the
methods of classical statistical physics were successfully explored in
economics and finance. See, e.g. \cite{Fi1}--\cite{Fi3} for quantum-like
financial models. We also remark that recently the methods developed for
non-Archimedean physical models which are widely used in string theory,
cosmology, spin glasses, e.g. \cite{pp1}, \cite{pp2}, started to be actively
applied in cognitive psychology e.g., \cite{p1}, \cite{p2}.

\section{Elements of quantum theory explored in the ASE model}

\label{PHYS}

We plan to explore the quantum laser model to describe mathematically the
possibility of \textit{Action Amplification by Stimulated Emission of Social
Energy} (ASE). We need not go deeply in the details of the quantum formalism
to present its features which will be explored in this paper (see, e.g. \cite%
{HB} for a non-physicist friendly introduction to the quantum formalism).
The main feature is \textit{discreteness} (\textquotedblleft
quantumness\textquotedblright ): the existence of stationary states of an
atom corresponding to discrete levels of energy; then spontaneous and
especially stimulated emissions of radiation by atoms. The basics of this
theory were set out by Einstein \cite{EN}.

Consider for simplicity the two level atom, it has the ground state $\psi
_{0}$ and the excited state $\psi _{\mathrm{ex}}$ corresponding to the
energy levels $E_{0}$ and $E_{1},$ respectively. The main point is that the
atom sufficiently sharply keeps one of those two states (at least ideally%
\footnote{%
The real situation is essentially more complicated than it is typically
described in textbooks on quantum mechanics. The most natural picture of the
energy distribution is given by two Gaussian distributions sharply
concentrated near their means, $E_{0}$ and $E_{1}.$ Thus their dispersions
are very small, but in reality they are nonzero. This remark is very
important for social applications. Here reality is even further from the
ideal model in physics.}).

The atom cannot be for ever in the state $\psi _{\mathrm{ex}};$ it has a
tendency to emit a photon and fall to the ground state $\psi _{0}.$ This
process is called \textit{spontaneous emission of radiation.} The crucial
characteristics of this process is that the energy of an emitted quantum
(nowadays known as photon) equals the difference between the energies of
levels: 
\begin{equation}
\Delta E=E_{1}-E_{0}.  \label{delta}
\end{equation}%
Thus, the fixed type of atoms (characterized by their energy levels) can
emit only photons of the fixed energy (the real situation is again more
complex and we again have to proceed with the Gaussian distribution with
mean value $\Delta E$). This is the origin of spectral lines which can be
observed experimentally (in reality these are Gaussian dimmed stripes).

However, different atoms in a population emit photons spontaneously in
different directions and at random moments of time. Such a type of emission
is characteristic for fluorescence and thermal emissions, see e.g., \cite{B1}%
. There is no \textit{coherence} in emission. The same relation (\ref{delta}%
) plays a key role in the absorption of energy by atoms. An atom in the
ground state can absorb only a photon of the energy $\Delta E.$ Photons with
energies different from this quantity are `ignored' by atoms of this type.
Even in the absence of external radiation sources, the atom can neither be
forever in the ground state: it jumps to the excited state (with some
probability). This is a consequence of vacuum fluctuations or in the
semiclassical models of the presence of the random background field. As was
remarked, the atom reacts only to background photons of the energy $\Delta
E. $

This story was about spontaneous quantum processes. Their analogs will not
play an essential role in the upcoming quantum-like social model. The main
role will be played by stimulated emission and absorption. The
inter-relation (\ref{delta}) gives the hint that if a population of atoms in
the ground state is subjected to the radiation composed of photons of energy 
$E_{\mathrm{ph}},$ then these atoms are able to absorb photons (with some
probability) only if $E_{\mathrm{ph}}=\Delta E,$ where the latter is
determined by (\ref{delta}). This is confirmed by quantum theory. This is 
\textit{stimulated absorption.} In the same way, if a population of atoms in
the excited state is subjected to the radiation composed of photons of
energy $E_{\mathrm{ph}},$ then these atoms emit photons (with some
probability) if 
\begin{equation*}
E_{\mathrm{ph}}=\Delta E.
\end{equation*}%
Here a `stimulated atom'\ does not absorb the `stimulating photon'. The atom
relaxes to the ground state and two photons are in flight. This is \textit{%
stimulated emission.} Thus, if an external photon with the energy $E_{%
\mathrm{ph}}=\Delta E$ stimulates emission from some atom then it results in
two photons of this energy. These two photons can stimulate emission from
two atoms, resulting in four photons and so on. There the number of emitted
photons increases exponentially.

The main distinguishing feature of this process which will play the
fundamental role in the upcoming social modeling is that this emission (in
opposition to spontaneous emission) generates the coherent beam of photons.
The emitted photon is a copy of the photon which had stimulated emission. In
particular, an atom emits a photon in the same direction as the light
passing by. It provides a beam which is sharply concentrated in one fixed
direction.

The coherence in a beam is not reduced to the spatial dimension: there can
be plenty of synchronization in this beam. In our social modelling
applications, we explore a possibility of such synchronizations. In the wave
picture, the main occurrence of coherence resides in the \textit{%
constructive and destructive interference}. Thus, the contributions of
different photons can be amplified (and very strongly) or canceled
(practically completely).

We finalize the discussion with the following list of quantum features:

\begin{enumerate}
\item Discrete levels of energy (for atoms and fields)

\item Bose-Einstein statistics of field quanta

\item Spontaneous emission

\item Stimulated absorption and emission

\item Coherent emission
\end{enumerate}

\section{Laser: light amplification by stimulated emission of radiation}

The quantum effects of stimulated emission and absorption were established
at the very beginning of quantum theory. However, only in the 1950's were
these effects realized in devices which are known nowadays as lasers.
Schematically, the laser has a simple structure. The \textit{gain medium} is
a population of atoms (with an identical structure of energy levels)%
\footnote{%
Impurities would contribute to decoherence of the emitted beam.} which are
excited by an external source of energy (pump). A pump based on a light
source, or an electrical field supplies energy for atoms to absorb and be
transformed into their excited states. Initially the majority of atoms in a
population are in the ground state, the minimum energy state. When the
number of particles in the excited state exceeds the number of particles in
the ground state (as the result of the pump), it is said that \textit{%
population inversion} is achieved. Then, for such population, \textit{the
amount of stimulated emission due to light that passes through is larger
than the amount of absorption.}

Hence the light sent to such population will be \textit{amplified} and the
output will be coherent. This process has only two components:

\begin{itemize}
\item The pumping of energy to the gain medium, the atom population, to
approach population inversion

\item Stimulated emission of light
\end{itemize}

For some types of lasers, this two component process leads to the required
amplification of light. And, for a moment, we restrict consideration to such
lasers. In other types of lasers, the beam obtained as the result of
stimulated emission is reflected from a mirror ($M1$) and send back through
the gain medium, again amplified, and reflected from another mirror ($M2$)$,$
and so on. This amplification process can be repeated a few times generating
higher amplification. However, as we pointed out, we proceed with the
simplest type of lasers combining pumping with stimulated emission.

We also remark that two level atoms are not the best gain medium: atoms with
a more complex level structure are used to produce better lasers. In the
quantum optics framework, the population inversion can be approached only in
a gain medium consisting of atoms having at least three levels and with a
special structure of transition probabilities. For our further studies, it
is important to remark that this is a consequence of coincidence of
Einstein's $B$-coefficients describing the transition probabilities for
stimulated absorption and emission, $B_{12}=B_{21}.$ This coincidence is
questionable in our quantum-like social studies. The field of social
information excitations is a boson field, i.e. its quanta satisfy the
Bose-Einstein statistics. However, there are no reasons to identify its
mathematical structure precisely with the electromagnetic field (although
the latter is convenient as giving the simplest model). Moreover, the
standard derivation of coincidence of Einstein's $B$-coefficients (for the
quantum electromagnetic field) is based on the assumption of approaching the
thermodynamical equilibrium and the probability distribution for the field's
energy described by the Planck law for black-body radiation. Even the
approachability of such an equilibrium in social modeling can be questioned.
Thus, in principle the social analogs of lasers based on two level systems
are possible. However, we ignore these technicalities (which are in fact
very important even in the social engineering of ASE): we want to present
just the basic scheme of amplification of coherent social information
excitations.{}

That is all: we need nothing more from quantum physics. We shall now
establish the correspondence between elements of the quantum physical and
the quantum-like social model.

\section{From \textquotedblleft it from bit\textquotedblright\ to the
quantum-like formalization of social information excitations}

\label{SQ}

In modern physics the purely information interpretation of physical laws
plays the important role apotheosized in Wheeler's \textquotedblleft it from
bit\textquotedblright\ \cite{WL}. D. Chalmers \cite{CL} summarised Wheeler's
views as follows:

{\small ``Wheeler (1990) has suggested that information is fundamental to
the physics of the universe. According to this ``it from bit'' doctrine, the
laws of physics can be cast in terms of information, postulating different
states that give rise to different effects without actually saying what
those states are. It is only their position in an information space that
counts.''}

The information approach in physics is very supportive to applications of
physical formalisms to the cognitive and social sciences. In particular,
Chalmers continued:

{\small ``If so, then information is a natural candidate to also play a role
in a fundamental theory of consciousness. We are led to a conception of the
world on which information is truly fundamental, and on which it has two
basic aspects, corresponding to the physical and the phenomenal features of
the world.''}

Recently the information approach to physics culminated in a variety of
information interpretations of quantum theory. We mention just a few of them:

\begin{enumerate}
\item Zeilinger-Brukner: quantum state as a presentation of (private)
information about possible results of measurements on a system \cite{Z1}-%
\cite{BZ};

\item Fuchs (in cooperation with Mermin, Caves and Schack), QBism, Quantum
Baeysianism: quantum state as presentation of subjective probabilities about
possible results of measurements on a system \cite{F1}- \cite{F3};

\item d'Ariano (in cooperation with Chiribella and Perinotti): derivation of
the quantum formalism from a set of information-theoretical postulates \cite%
{DA1}--\cite{DA3}.
\end{enumerate}

In the information approach, quantum mechanics is not about a `quantum
world', but about our (observers) predictions on the possible results of
measurements which can be performed on micro-systems. This viewpoint is
close to the original views of Bohr and especially Pauli, see \cite{PLKHR}.
Of course, the purely informational interpretation of quantum mechanics and
physics in general does not deny reality. For example, Bohr never denied the
existence of atoms as material entities. However, only the state structure
of atoms is described by the quantum formalism (whether a deeper description
is possible is still the subject of stormy debates in quantum foundations).
From this viewpoint, any entity whose state structure can be mapped onto the
state structure of atoms can be in principle described by the same quantum
formalism. By using the information interpretation of quantum field theory
we can view quantum fields as \textit{quantized information fields.} Their
quanta, excitations of quantum fields, can be interpreted as quanta of
information. In particular, the quantum electromagnetic field can be treated
as a special information field with quanta known as photons. We remark that
the spatial wave-function of a photon is not well defined. Therefore it
cannot be interpreted as a localized physical particle and it cannot be
interpreted as a physical wave. Thus the most consistent way is to treat it
as a quantum of information, given by the momentum and polarization vectors.

We now explore the information viewpoint to physical formalisms and borrow
them for our social modeling. Individuals are mapped to atoms: we can speak
about \textit{\textquotedblleft social atoms\textquotedblright }, $s$-atoms.
Human populations, societies, are mapped to atom populations. In particular,
in our model human societies play the role of gain mediums. The information
exchange between $s$-atoms is formally modeled with the aid of a quantized
information field. Its quanta are interpreted as social information
excitations. It is natural to model the information field as a boson field
(see energy-considerations below). As was remarked above, the simplest (from
the mathematical viewpoint) boson field is the electromagnetic field.
Therefore, we proceed with the information field (transmitting the
information to and from $s$-atoms) which is described as the quantum
electromagnetic field (this is just borrowing from physics the concrete
model of information exchange, nothing more). Thus, the quanta of
information carrying social excitations are modeled as social analogs of
photons, $s$-photons (see section \ref{SPH} for further discussions).

One of the basic assumptions of our model is that the states of $s$-atoms
and photons can be characterized by a quantity which can be transferred into
social activity (`work') of individuals or groups of individuals. We call it
the \textit{social energy.} As well as in physics, the social energy is a
primary quantity which cannot be derived from more elementary ones. Again,
as in physics, this is simply a tool for the quantitative characterization
of possible activities of individuals\footnote{%
In physics a better understanding of the features of energy is approached
through the description of mutual transformations of various types of
energy. In thermodynamical studies, see \cite{HM}, \cite{HM1} for a similar
attempt for the `information energy'\ -- the energy of expectations of
traders of the financial market.}. On the operational level, in quantum-like
models the social energy is represented by an operator (Hamiltonian)
generating the dynamics of a mental state, similarly to quantum mechanics.
This quantity was successfully used in quantum-like financial models
representing the energy of expectations of traders \cite{QL4}. We remark
that the value of the social energy does not determine the concrete
structure of an excitation and a possible action induced by it (in the same
way as in physics the value of energy of an excited state does not determine
the direction in which the photon can be emitted). We understood well that
the problem of the interpretation of the social energy has to be analyzed in
more detail. We cannot do this in the present paper, but see section \ref%
{SPH1} for a brief comparison of the notions of the quantum and social
energies.

The next fundamental assumption is that, for some societies, \textit{the
levels of the social energy for $s$-atoms are quantized}, sharply
concentrated. For example, the ground state (the minimal social energy in
this society) and the state of a social excitement. In the simplest model,
as in the previous sections, we characterize these states by just two
numbers $E_{0}$ and $E_{1}.$ In reality they can be collections of
parameters characterizing states. We remark that the sharpness of the levels
of social energy is dimmed: it is of the Gaussian type, cfr. section \ref%
{PHYS}. We also assume that the information fields are quantized.

We now motivate that the basic energy absorption-emission relation (\ref%
{delta}) holds even for social energy. We proceed under the assumption of
the discreteness of energy levels of individuals (in a population under
study) and the quantization of information fields, namely, the transmission
of social excitations by quanta of information, see section \ref{SPH} for a
foundational discussion. Since a few quanta of the information field
(realized, e.g. as TV-communications) can carry the same social energy,
their distributions have to obey the laws of Bose-Einstein statistics, i.e.
information fields are boson fields. And, as a model, we select the simplest
of them, namely, the quantum electromagnetic field.

In our framework a social analog of the following property of the physical
photon is crucial : an atom cannot `eat'\ a part of photon: it either eats
the whole portion of energy carried by the photon or simply ignores this
portion, if its energy is too small or too large to match with the energy
structure of the atom. This property matches well with the absorption of
information by humans: an individual typically does not try to split a
communication, e.g. in TV-news, into pieces and takes into account some
concrete piece of it. The whole communication is either `eaten'\ or not.

If the $s$-atom has only one excited state, then automatically it can `eat'\
only a communication carrying the social energy given by the social analog
of the formula (\ref{delta}). Thus the validity of this formula in social
processes is a consequence of the discreteness of the energy levels of the
structured human media: the discreteness of energy carried by information
communications; the tendency of humans to absorb communications as
indivisible entities, quanta. And the discreteness implies that only
information quanta of concrete energy can be absorbed by individuals%
\footnote{%
Heuristically this picture is very natural. Humans suppress, e.g. the
information communications carrying the energy which is essentially higher
than the level of socially acceptable excitation in a population with
structured social energy. We all ignore the communication that, since the
1970's around 50\% of living species disappeared from Earth. This excitation
carries a too high energy. An individual in an energy structured population
is not ready to process such an excitation (at least not consciously).}.

We presented the model of absorption of social radiation. Now if the $E_{1}$%
-excited individual relaxes, approaching the ground state, she/he can emit
only a social information excitation potentially leading to a social action,
having the energy given by (\ref{delta}), since she/he could not relax to
some level in between $E_{0}$ and $E_{1}.$ This emission can be spontaneous:
an individual cannot be forever in the excited state: she/he relaxes to the
ground state. Spontaneous as in physics means a-causal. She/he relaxes
without any definite cause for falling into the ground state. It is
impossible to predict when and in which way, a human relaxes from the state
of excitement and `emits a social information excitation'. The latter may
lead to a social action of the corresponding social energy. However, even in
physics many photons disappear in a medium and noisy background radiation,
in the same way as many social information excitations, which are
excitations in the information space, do not lead to real actions. They
disappear in a noisy information background. Such spontaneous mental
relaxations definitely match with human behavior.

However, we are more interested in a social analog of the stimulated
emission. This is the most complicated part of the model. In quantum
physics, to derive the stimulated emission one has to explore the wave
picture of the photon and the coupling between the photon's frequency and
energy. For excitations of a social quantum-like field, the proper frequency
interpretation is not so straightforward as it is for photons (see section %
\ref{FE} for a discussion). For a moment, we discuss the stimulated emission
of social information excitations from the heuristic viewpoint (cfr. with
Einstein's derivation in the pioneering work \cite{EN}). An $s$-atom $A$ in
the excited state interacts with an information field. The latter is
quantized, it is composed of excitations of various energies and directed to
various social actions. The stimulated emission means that $A$ ignores all
information quanta, communications, having the energy different from $\Delta
E=E_{1}-E_{0}$ (again in reality we have a Gaussian distribution with the
mean value $\Delta E).$ However, if a communication has the same energy%
\footnote{%
For example, in modern Western society, this is a web-call for an
anti-globalist demonstration and not a call for a military operation against
the government.} as $\Delta E$ then the probability of $A$'s relaxation
increases essentially and if $A$ relaxes, then her/his possible action is
identical to the potential action carried by the stimulating excitation%
\footnote{%
The latter is natural: by accepting the communication about the
anti-globalist demonstration one will go to such a demonstration and not to
a demonstration against the discrimination of women, even if both carry the
social energy of the same degree.}.

\section{The structuring of social energy}

The primary assumption for the possibility of ASE is the discrete structure
of the social energy for individuals in some human societies. In a society
where individuals have a continuous spectrum of the social energy, ASE is
impossible.

In any society an individual can become mentally excited to some degree
(with some probability). For example, consider the various degrees of states
of a social protest: from carrying an opposition mentality to participating
in demonstrations, barricades and revolutions. If a society is clustered
into a variety of groups of various degrees of excitement, this is not the
case in our modeling. Consider a society where for example one individual is
prepared just for demonstration and another individual wants only to express
oppositional views in front of his wife or friend, whilst another individual
is ready to go to protest against the system, but in a peaceful way. There
may be also groups planning actions of different degrees of violence. Such
type of society is difficult to subject to a stimulated \textit{coherent
excitation.}

The degree of excitement has to be homogeneous, i.e. it is structured in
such a way that it is `natural'\ to belong to the same level of excitement
or to be in the ground state (we remind the reader that the simplest two
level model is under consideration).

In modern Western society it is natural and well accepted to have social
excitement at a concrete level: e.g. to demonstrate against cuts to
education or social needs (the energy level $E_{1}),$ but not to be excited
about a revolution against the system. We remind ourselves, that we do not
speak about sharp levels, but about Gaussian distributions concentrated
around the levels. The number of people with social energy $E>>E_{1}$ is
negligibly small. There are practically no people who are ready to struggle
to destroy the system. At the same time, not so many people have the energy
distributed in the gap $\Delta E=E_{1}-E_{0}.$ People are either `socially
active', of the same $E_{1}$-level or simply socially passive, of the same $%
E_{0}$-level. This society is well structured and it can serve as the basis
for ASE (please see section \ref{CONCL}).

At the beginning of the 20th century, Russian society was well structured,
but with essentially higher energy $E_{1}.$ It was very fashionable (even
for intellectuals and a part of bourgeoisie) to be in the state of
revolutionary excitement. People were not interested in social activities
with energies distributed in the gap $\Delta E=E_{1}-E_{0}.$ Of course, such
groups existed, as, e.g. the followers of Leo Tolstoy, the Tolstoyan
movement, but they were negligibly small compare to the total socially
active population. From the quantum viewpoint these are just impurities in
the gain medium.

The energy structure of a society is determined by the social context which
combines cultural, historical, economic, financial, political and even
weather conditions\footnote{%
In general the notion of context plays a crucial role to motivate
applications of the operational quantum formalism to cognition, psychology
and social science. Cognition (both on individual and collective levels) is
intrinsically contextual, and so are quantum phenomena. Here we have no
possibility to discuss the issue of contextuality in more detail, see, e.g., 
\cite{QL1}, \cite{G0}--\cite{D2}.}. Since social energy is an informational
quantity, the aforementioned components of context are also of a pure
informational nature. Roughly speaking, it is not the real political
situation, but its representation by various sources of information which is
important. Nowadays, the mass-media and internet are the main sources of
discretization of social energy. An individual feels comfortable to belong
either to the `socially active part of society' (the same level $E_{1}$) or
to live an `ordinary life'. Internet-communities play an important role in
the energy structuring of the human gain medium and in the homogenization of
the excitation strength. However, in this paper we have no possibility for a
detailed analysis of this psychological phenomenon.

\section{Action Amplification by Stimulated Emission of Social Energy}

The social laser is based on the human gain medium, a population with a
discrete structure of levels of social energy. We proceed with the simplest
model: a population with the two level structure\footnote{%
In reality, as in laser physics, more complex structures of energy levels
have to be explored. We shall consider such models in upcoming publications.}%
. Typically, a majority of people are being in the ground state, the state
of the minimal social energy. However, by pumping social energy into a
population its inversion can be approached, so a majority of people will be
excited. In physics, pumping typically is of the short pulse-form. A strong
pulse delivers a lot of energy to the gain medium, but it has to be short
otherwise it may destroy the gain medium. The social energy pumping has to
have the similar structure: a short pulse of news. When population inversion
is achieved, this is the time to start stimulated emission. The human gain
medium is exposed to the flow of coherent news, having the energy given by (%
\ref{delta}). Stimulated emission starts and it induces a cascade of
coherent social information excitations (of the exponentially increasing
strength) leading to a coherent action of this population, e.g. in the form
of a social protest. The latter can imply the realization of huge social
energy.

\section{Interpretational issues}

\label{INT}

\subsection{Quantization of human excitations}

\label{SPH}

The main interpretation problem in using the photon metaphor for the
mathematical modeling of human excitations is that even in quantum physics
the notion of photon is the subject of intense debate (since the invention
of the light-quantum by Einstein in 1905). Roughly speaking, the following
problem has been debated during the last one hundred years: does an
electromagnetic field quantize in a vacuum? Does a photon `exist'\ in the
absence of matter? Is the notion of photon meaningful only for the
description of the process of interaction of the electromagnetic field with
matter? Modern quantum physics is based on Einstein's viewpoint: yes, a
photon exists even in the absence of matter\footnote{%
However, we remark that some fathers of quantum mechanics, e.g. Lamb and
Lande disagreed with Einstein (and Lamb died not long ago, he preserved his
viewpoint in the light of all the successes of quantum mechanics), see \cite%
{Beyond} for details and modern attempts to proceed in the Lamb-Lande
direction. The main argument for the interpretation of a photon as simply an
excitation, an action transmitter, is that the presence of photons can be
detected only with the aid of material particles, in the process of
detection. We can detect only their actions and typically the act of
detection (the action's realization) leads to the photon destruction. It is
worse to mention that the position representation of photons (representation
in space-time) is not well-posed, the wave function of photon cannot be
properly defined (although the last hundred years were characterized by
numerous attempts to proceed in this direction). As a result of this
problem, the photon is typically treated not really as a particle, similar
to e.g. an electron, but as an \textit{excitation} of the quantum field.}.
In our framework the question of the `existence of mental photons'\ can be
formulated as follows: can human actions (or more precisely excitations to
perform actions) be treated independently from individuals? Do human
excitations live their own life? This is an interesting philosophical and
methodological question. However, for the moment we ignore it completely (so
as to avoid being involved in a debate similar to the debate on photon
existence).

\subsection{Social versus quantum energy}

\label{SPH1} Now we briefly discuss and compare the notions of energy in
quantum physics and quantum-like sociology. The reader might feel that the
notion of social energy invented in section \ref{SQ} is fuzzy and less
intuitive than the notion of physical energy. He/she is correct with respect
to the comparison of \textit{classical physical energy} with social energy.
However, the notion of energy in quantum mechanics is less intuitive than in
classical physics. One could not simply assign a concrete value of energy to
a quantum system, if it is not in a stationary state (e.g. an atom can be in
a superposition of the ground and excited states, i.e. its energy is neither 
$E_{0}$ nor $E_{1}$ and at the same time both $E_{0}$ and $E_{1})$. The
concrete value of energy is determined only as the result of its realization
in the process of detection. In the same way in general the individual's
state cannot be characterized by the fixed value of the social energy.
Stationary states (at least different from the ground state, the state of
the minimum of social energy) are not stable.

\subsection{Social versus photon's momentum and polarization}

THe photon's state can be characterized by the momentum vector $p$ and
polarization vector $s.$ The first one represents operationally the
direction of propagation of the photon and the second represents a special
internal degree of freedom of the photon. By using the same mathematical
model we equip the $s$-photon with a (social) momentum and polarization. The
first specifies the general `direction'\ of a possible social action, e.g.
anti-war activity, anti-globalism etc...The second represents concrete
characteristics of the social excitation `directed'\ by the momentum $p.$
For example, $k=$`anti-war in Vietnam activity', $s=$`March 25-26 (1966).
Days of International Protest. Organized by the National Coordinating
Committee to End the War in Vietnam.'\ In quantum mechanics the polarization
space has the dimension two. However, in principle we are not rigidly
coupled to the photon model. For us, the photon is just one of the possible
transmitters of action. The quantum theory of \textit{gauge fields} provides
us with plenty of mathematical models with more complex spaces of internal
degrees of freedom.

In previous considerations we discussed only the interpretation of the
direction encoded in the momentum, i.e., given by the normalized vector $%
\frac{p}{\vert p\vert}.$ In physics its length is proportional to the photon
energy. We can proceed in the same way (here it will be the definition of
the magnitude of the social momentum).

\subsection{Frequency interpretation}

\label{FE}

By interpreting the $s$-photon as a `quantum of possible action'\ and
assuming (by extending Einstein's idea to the social domain) that the social
energy is quantized even in the absence of interaction with concrete
individuals, we can treat the social information space as filled by quanta
of possible actions. Such $s$-photons are purely information entities. How
can we characterize their energy (before its realization)? It seems that,
for such a purely information quantity, its energy can be characterized by
the \textit{frequency} $\nu $ of its appearance in the information space
(e.g. in TV-news, in newspapers and Internet). Thus, it is natural to couple
the frequency of communications with the energy of a social quantum of
excitation. What is the form of the frequency-energy relation? In quantum
physics this relation is given by the Einstein formula: 
\begin{equation}
E_{\mathrm{ph}}=h\nu ,  \label{EF}
\end{equation}%
where $h$ is the Planck constant. This is the simplest possible law, the
linear one. One may try to keep this relation even in social modeling. In
quantum physics (\ref{EF}) is treated as the explicit relation. In our study
we can treat its social analog as just a linear approximation of a more
complex nonlinear law. The presence of the fixed constant $h$ in (\ref{EF})
is a delicate problem. We definitely cannot expect that a kind of `social
Planck constant' exists. It is more natural to expect that different types
of social excitations are characterized by coefficients of proportionality
of different magnitudes (cfr. for discussions on a financial analog of the
elementary quantum of action, see \cite{QL1}, \cite{QL4}).

The reader has already noticed that in the model of the social laser we
proceeded without the frequency interpretation for the $s$-photon. It is
clear why. The frequency is coupled to wave features of the photon. In
physics the basic law of radiation is given by the Planck-Einstein formula: 
\begin{equation}
\Delta E=E_{1}-E_{0}=h\nu .  \label{delta_X}
\end{equation}%
We split it: only its first part, see (\ref{delta}), was in play. Now we
discuss its second part, see (\ref{EF}).

Of course, even a physical photon (quantum of the electromagnetic field)
cannot be simply imagined as a classical wave propagating in space modeled
as $\mathbb{R}^{3}.$ However, such a heuristic picture has at least some
illustrative power. For the $s$-photon, $\mathbb{R}^{3}$ is used to encode
the `directions'\ of action (this is simply a linear space representation of
mental states which is widely used in cognitive science, psychology and the
social and political sciences, e.g. \cite{VG}.) Even heuristically it is
difficult to operate with waves in this action-space. However, we should not
forget that operationally the only exhibition of (physical) photon wave
features is the interference of probabilities of detections. Similar
interference features of mental entities have been studied sufficiently well 
\cite{QL1}, \cite{QL2}.

\section{Concluding remarks}

\label{CONCL} By exploring the quantum principles of laser functioning we
formulated the corresponding principles of functioning of the `social
laser', generating Action Amplification by Stimulated Emission of Social
Energy (ASE). The analogs of fundamental quantum principles leading to the
possibility of the creation of social lasers can be formulated as follows:

\begin{itemize}
\item the social energy of individuals in some human populations, `gain
mediums', can be structured in discrete levels;

\item a human gain medium can absorb and emit information excitations only
with energies equal to the difference between energies of discrete levels,
see (\ref{delta});

\item an information excitation having the energy matching the discrete
levels of an individual, stimulates emission of an excitation in the state
which is identical to the state of the stimulating excitation.
\end{itemize}

The structure of the social laser is similar to the structure of the
physical one:

\begin{itemize}
\item a human gain medium;

\item pumping of social energy to it -- to approach population inversion;

\item stimulating of emission from the gain medium.
\end{itemize}

In this paper we presented a general quantum-like model of ASE. We do not
try to couple it to concrete social protests, including the recent ones
mentioned in the introduction. This is not our task. The descriptions of
special human gain mediums, the structures of their discrete levels of
social energy and the machinery of energy pumping and stimulated emission
can be done by experts in the social and political sciences.

Finally we remark that the expression `stimulated emission of social
information excitations'\ might be misleading. One needs not to imagine
stimulation as the process of consciously designed stimulation of a human
population (after approaching the population inversion state) by a coherent
flow of excitations. In physics lasers are merely known as artificially
designed devices. However, nature creates lasers by itself without the
conscious design of physicists. `Natural lasers'\ are well known in astronomy%
\footnote{%
The first `natural'\ laser in space was detected by scientists on board
NASA's Kuiper Airborne Observatory as they trained the aircraft's infrared
telescope on a young, very hot, luminous star in the constellation Cygnus,
see www.nasa.gov/home/hqnews/1995/95-148.txt}. In a same way, human (and
other biological) societies are able to create `natural social lasers',
i.e., ASE self-generated by human societies. It seems that the majority of
ASE during the last years are of a natural origin. The modern information
societies can produce `natural social lasers'\ as the result of the creation
of extremely powerful communication channels, especially the Internet. We
have already discussed the role of the mass media and the Internet in the
discrete structuring of the social energy. They also produce periodically
strong information pulses pumping social energy into the population. Such
pumping excites extended layers of the population and leads to population
inversion. Even the stimulated emission of social information excitations
need not be planned and designed consciously. Coherent news (flows of
identical social information excitations) can be produced without conscious
design, simply as the result of the homogeneity of information flows
delivered by the mass-media and the Internet.

We point out that one of the surprising features of the recent social
protests and revolutions, is the absence of well formulated political
programs and strong political leaders, see I. Krastev \cite{P4}. This
feature was widely and controversially debated in political studies, but
without coming to a consensus on its meaning and origin. However, it matches
well with the functioning of natural social lasers. Here, if a social group
approached the state of population inversion, then any coherent flow of news
of the energy matching with the energy level structure of this group can
generate ASE. There is no need in writing say the `Manifesto of the
Communist Party' (issued by Marx in 1848); modern ASE happens without such
figures as Martin Luther, Karl Marx, Vladimir Lenin,.... In any event if the
hypothesis that the modern information society can self-generate ASE is
correct, then in the future one can expect the increase of the frequency of
ASEs throughout the world - simply as one of the natural features of the
modern information society\footnote{%
At the same time, we understand that the connection of recent social
processes with a `natural social laser' can be an illusion. If some powerful
mechanisms can arise in a natural way, then very clever people will use and
optimize them. We repeat that the analysis and interpretation of these
events in the social and political literature are very controversial \cite%
{P0}--\cite{P4}. There are claims that the absence of political programs and
strong political leaders is an illusion; these programs and leaders are just
hidden. }.

Finally, when considering the possibility of the application of ASE to model
social protests we make the following remark. Population inversion means
that more than half of the population is excited. In fact, the real actions
emanating out of protests involve a minority of population. Here it is
important to distinguish the emission of a coherent wave of $s$-photons and
the realization of their energy in real social actions. ASE (as well as
stimulated emission of energy in physics) describes only the emission of the
quanta of energy. The real social actions can be treated as analogs of the
measurements performed on photons, i.e. the interaction of the field-quanta
with atoms. In our social laser model, the majority of the population emits
coherent $s$-photons, but only a fraction of them is `detected', e.g. in
clashes with police and army. However, the presence of a strong coherent
wave of opposition plays a crucial role, at least in the aforementioned
recent social protests and revolutions. In fact, the presence of such an
information wave restricts the force of reactions from governmental organs.

\section*{Acknowledgment}

This paper was partially supported by the grant ``Mathematical modeling of
complex hierarchic systems'' of Linnaeus University and the visiting
professor fellowship at the Center for Quantum BioInformatics (QBIC), Tokyo
University of Science (October 2014). The author would like to thank I.
Basieva for numerous discussions on principles of laser-functioning.

\end{document}